# An Introduction to Fluorescence Resonance Energy Transfer (FRET)

Syed Arshad Hussain
Department of Physics, Tripura University, Suryamaninagar-799130, Tripura, India

**Abstract:** Recent advances in Fluorescence Resonance Energy Transfer (FRET) provides a way to measure and understand different biological systems and molecular interactions in nanometer order. In this report the introduction and principle of the FRET process have been explained.

## Introduction:

The Fluorescence Resonance Energy Transfer (FRET) between two molecules is an important physical phenomenon with considerable interest for the understanding of some biological systems and with potential applications in optoelectronic and thin film device development [1, 2]. The technique of FRET, when applied to optical microscopy, permits to determine the approach between two molecules within several nanometers. FRET was first described over 50 years ago, that is being used more and more in biomedical research and drug discovery today. FRET is a distance dependant radiationless transfer of energy from an excited donor fluorophore to a suitable acceptor fluorophore, is one of few tools available for measuring nanometer scale distances and the changes in distances, both in vitro and in vivo. Due to its sensitivity to distance, FRET has been used to investigate molecular level interactions. Recent advances in the technique have led to qualitative and quantitative improvements, including increased spatial resolution, distance range and sensitivity.

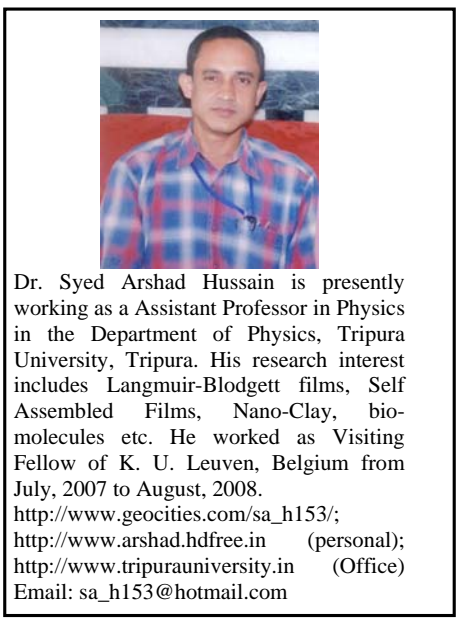

Dr. Syed Arshad Hussain is presently working as a Assistant Professor in Physics in the Department of Physics, Tripura University, Tripura. His research interest includes Langmuir-Blodgett films, Self Assembled Films, Nano-Clay, bio-molecules etc. He worked as Visiting Fellow of K. U. Leuven, Belgium from July, 2007 to August, 2008.
http://www.geocities.com/sa_h153/;
http://www.arshad.hdfree.in    (personal);
http://www.tripurauniversity.in    (Office)
Email: sa_h153@hotmail.com

The mechanism of fluorescence resonance energy transfer involves a donor fluorophore in an excited electronic state, which may transfer its excitation energy to a nearby acceptor chromophore in a non-radiative fashion through long-range dipole-dipole interactions. The theory supporting energy transfer is based on the concept of treating an excited fluorophore as an oscillating dipole that can undergo an energy exchange with a second dipole having a similar resonance frequency. In this regard, resonance energy transfer is analogous to the behavior of coupled oscillators, such as a pair of tuning forks vibrating at the same frequency. In contrast, radiative energy transfer requires emission and reabsorption of a photon and depends on the physical dimensions and optical properties of the specimen, as well as the geometry of the container and the wavefront pathways. Unlike radiative mechanisms, resonance energy transfer can yield a significant amount of structural information concerning the donor-acceptor pair.

Resonance energy transfer is not sensitive to the surrounding solvent shell of a fluorophore, and thus, produces molecular information unique to that revealed by solvent-dependent events, such as fluorescence quenching, excited-state reactions, solvent relaxation, or anisotropic measurements. The major solvent impact on fluorophores involved in resonance energy transfer is the effect on spectral properties of the donor and acceptor. Non-radiative energy transfer occurs over much longer distances than short-range solvent effects, and the dielectric nature of constituents (solvent and host macromolecule) positioned between the involved fluorophores has very little influence on the efficacy of resonance energy transfer, which depends primarily on the distance between the donor and acceptor fluorophore.

A pair of molecules that interact in such a manner that FRET occurs is often referred to as a donor-acceptor pair. The phenomenon of FRET is not mediated by photon emission. Also it does not even require that the acceptor chromophore to be fluorescent. Although in most of the applications the donor and the acceptor are fluorescent.

**Principle of Fluorescence Resonance Energy Transfer (FRET):**

In the process of FRET, initially a donor fluorophore absorbs the energy due to the excitation of incident light and transfer the excitation energy to a nearby chromophore, the acceptor.

$$D + h\upsilon \rightarrow D^*$$
$$D^* + A \rightarrow D + A^* \quad [D \rightarrow \text{donor}, A \rightarrow \text{Acceptor}]$$
$$A^* \rightarrow A + h\upsilon'$$

Energy transfer manifests itself through decrease or quenching of the donor fluorescence and a reduction of excited state lifetime accompanied also by an increase in acceptor fluorescence intensity. Figure 1 is a Jablonski diagram that illustrate the coupled transitions involved between the donor emission and acceptor absorbance in FRET. In presence of suitable acceptor, the donor fluorophore can transfer its excited state energy directly to the acceptor without emitting a photon.

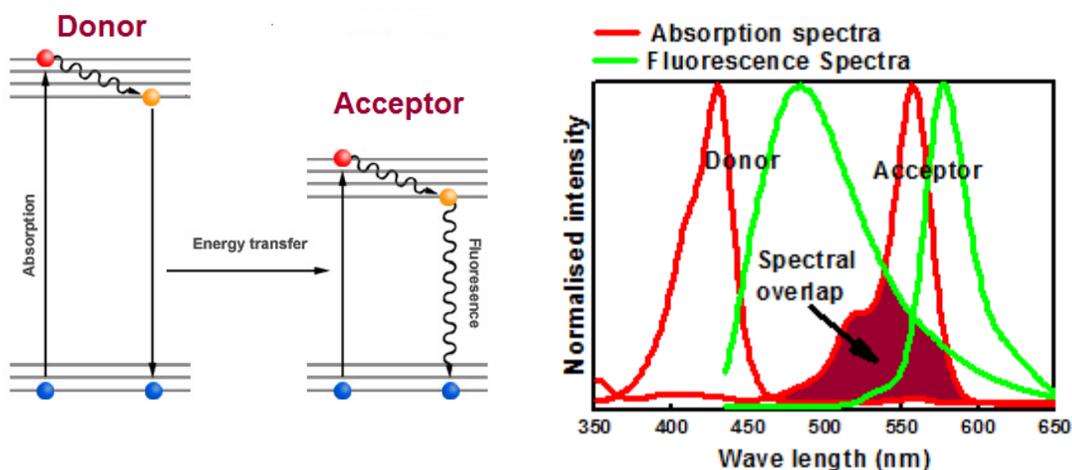

**Figure 1:** Jablonski diagram illustrating the FRET process.

**Figure 2:** Absorption and fluorescence spectra of an ideal donor-acceptor pair. Brown coloured region is the spectral overlap between the fluorescence spectrum of donor and absorption spectrum of acceptor.

There are few criteria that must be satisfied in order for FRET to occur. These are: (i) the fluorescence emission spectrum of the donor molecule must overlap the absorption or excitation spectrum of the acceptor chromophore (figure 2). The degree of overlap is referred to as spectral overlap integral (J). (ii) The two fluorophore (donor and acceptor) must be in the close proximity to one another (typically 1 to 10 nanometer). (iii) The transition dipole proentations of the donor and acceptor must be approximately parallel to each other. (iv) The fluorescence lifetime of the donor molecule must be of sufficient duration to allow the FRET to occur.

Förster [3] showed that the efficiency of the FRET process ($E_{FRET}$) depends on the inverse sixth power of the distance between the donor and acceptor pair (r) and is given by:

$$E_{FRET} = R_0^6 / (R_0^6 + r^6) \quad (1)$$

Where $R_0$ is the Förster radius at which half of the excitation energy of donor is transferred to the acceptor chromophore. Therefore Förster radius ($R_0$) is referred to as the dietance at which the efficiency of energy transfer is 50%.

The Förster radius ($R_0$) depends on the fluorescence quantum yield of the donor in the absence of acceptor ($f_d$), the refractive index of the solution ($\eta$), the dipole angular orientation of each molecule $(K^2)$ and the spectral overlap integral of the donor-acceptor pair $(J)$ and is given by [3],

$$R_0 = 9.78 \times 10^3 (\eta^{-4} . f_d . J)^{1/6} A^0 \qquad (2)$$

In summary, the rate of FRET depends upon the extent of spectral overlap between the donor-acceptor pair (figure 2), the quantum yield of the donor, the relative orientation of the donor-acceptor transition dipole moments and the distance separating the donor-acceptor chromophore. Any event or process that affects the distance between the donor-acceptor pair will affect the FRET rate, consequently allowing the phenomenon to be quantified, provided that the artifacts can be controlled or eliminated. As a result, FRET is often referred to as a `spectroscopic/molecular ruler´, for example to measure the distance between two active sites on a protein that have been labelled with suitable donor-acceptor chromophore, and therefore monitoring the conformational changes through the amount of FRET between the fluorophores.

**Detection of Fluorescence Resonance Energy Transfer (FRET):**

The detection and quantitaion of FRET can be made in a number of different ways. Simply the phenomenon can be observed by exciting a specimen containing both the donor and acceptor molecules with light emitted at wavelengths centered near the emission maximum of the acceptor. Because FRET can result in both a decrease in fluorescence of the donor molecule as well as an increase in fluorescence of the acceptor, a ratio metric determination of the two signals can be made. The advantage of this method is that a measure of interaction can be made that is independent of the absolute concentration of the sensor. Because not all acceptor moieties are fluorescent, they can be used as a means to quench fluorescence. In these instances, those interactions that result in a fluorescent donor molecule coming in close proximity to such a molecule would result in a loss of signal. Inversely, reactions that remove the proximity of a fluorescent donor and a quencher would result in an increase in fluorescence. Figure 3 illustrates the detection of FRET by observing the fluorescence spectra of the donor-acceptor pair.

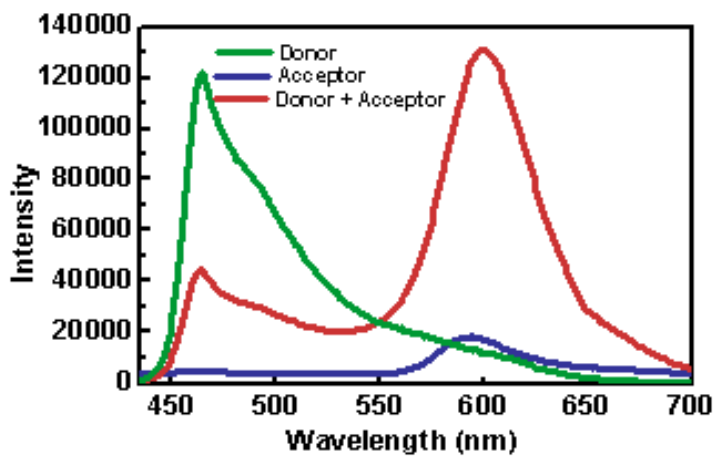

**Figure 3:** Fluorescence spectra of individual donor and acceptor and the mixed solution of donor and acceptor. Here the acceptor fluorescence increases by almost six times in presence of donor. Where as the donor fluorescence becomes one third of the individual donor fluorescence. This ratiometric change in fluorescence intensity is compelling visual evidence of FRET. The fluorescence spectra were recorded by exciting the emission maximum of the donor. [Here the fluorophore used are 3- octadecyl-2[3-octadecyl-2 (3H)-benzothizolidene) methyl] benzothiazolium perchlorate (donor), Octadecyl rhodamine B (acceptor)].

Another alternative method, growing popularity rapidly is to measure the fluorescence lifetime of the donor fluorophore in the presence and absence of the acceptor chromophore. FRET will cause a decrease in excited lifetime of the donor fluorophore.

**Application of Fluorescence Resonance Energy Transfer (FRET):**

The strong distance-dependence of the FRET efficiency has been widely utilized in studying the structure and dynamics of proteins and nucleic acids, in the detection and visualization of intermolecular association and in the development of intermolecular binding assays [4].

FRET is a particularly useful tool in molecular biology as the fraction, or efficiency, of energy that is transferred can be measured [5], and depends on the distance between the two fluorophores. The distance over which energy can be transferred is dependent on the spectral characteristics of the fluorophores, but is generally in the range 10–100A°. Thus, if fluorophores can be attached to known sites within molecules, measurement of the efficiency of energy transfer provides an ideal probe of inter- or intramolecular distances over macromolecular length scales. Indeed, fluorophores used for this purpose are often called ''probes''. Techniques for measuring FRET are becoming more sophisticated and accurate, making them suitable for a range of applications [5]. FRET has been used for measuring the structure [6-8], conformational changes [9] and interactions between molecules [10,11], and as a powerful indicator of biochemical events [12]. Further applications can be found in the reviews of Van der Meer et al. [13], Lakowicz [14], or Selvin [15]. Although the challenges of labeling molecules with fluorophores and making accurate measurements of the fluorescence emitted by them are being overcome, a number of difficulties still remain when examining real-life systems